\documentclass[manuscript,screen]{acmart}

\AtBeginDocument{%
  }

\title{``Don't Be Afraid, Just Learn'': Insights from Industry Practitioners to Prepare Software Engineers in the Age of Generative AI}%

\author{Daniel Otten}
\orcid{0009-0001-2702-0541}
\affiliation{%
 \institution{William \& Mary}
 \city{Williamsburg, VA}
 \country{USA}
}
\email{dsotten@wm.edu}

\author{Trevor Stalnaker}
\orcid{0009-0005-6000-4227}
\affiliation{%
 \institution{William \& Mary}
 \city{Williamsburg, VA}
 \country{USA}
}
\email{twstalnaker@wm.edu}

\author{Nathan Wintersgill}
\orcid{0009-0006-2123-7412}
\affiliation{%
 \institution{William \& Mary}
 \city{Williamsburg, VA}
 \country{USA}
}
\email{njwintersgill@wm.edu}

\author{Oscar Chaparro}
\orcid{0000-0003-2838-685X}
\affiliation{%
 \institution{William \& Mary}
 \city{Williamsburg, VA}
 \country{USA}
}
\email{oscarch@wm.edu}

\author{Denys Poshyvanyk}
\orcid{0000-0002-5626-7586}
\affiliation{%
 \institution{William \& Mary}
 \city{Williamsburg, VA}
 \country{USA}
}
\email{denys@cs.wm.edu}

\author{Douglas Schmidt}
\affiliation{%
 \institution{William \& Mary}
 \city{Williamsburg, VA}
 \country{USA}
}
\email{douglas.c.schmidt@wm.edu}

\date{January 2026}

\usepackage{ifthen}

\newboolean{showcomments}
\setboolean{showcomments}{true}

\newboolean{showboxes}
\setboolean{showboxes}{true}

\usepackage{verbatim}

\usepackage[utf8]{inputenc}
\usepackage{url}
\usepackage{colortbl}
\usepackage{balance}
\usepackage{xspace}
\usepackage{graphicx}
\usepackage{bold-extra}
\usepackage{paralist}
\usepackage{multirow}
\usepackage{listings}
\usepackage{boxedminipage}
\usepackage{soul}
\usepackage{enumitem}
\usepackage[normalem]{ulem}
\setlist{nolistsep,leftmargin=.5cm}
\usepackage{booktabs}
\usepackage{tabularx}
\usepackage{svg}
\usepackage{calc}
\usepackage{float}
\usepackage{multirow}
\usepackage[normalem]{ulem}
\useunder{\uline}{\ul}{}
\usepackage[most]{tcolorbox}
\usepackage{caption}
\usepackage{microtype} 
\usepackage{algorithmic}
\usepackage{textcomp}
\usepackage[dvipsnames]{xcolor}

\usepackage{wrapfig}
\usepackage{subcaption}
\usepackage{cleveref} 
\usepackage{adjustbox}
\usepackage{totcount}

\usepackage{tikz}
\usetikzlibrary{shapes.geometric, arrows.meta, positioning, decorations.pathmorphing}

\usepackage{breakurl}

\usepackage{hyperref}

\newcommand{\nb}[2]{}

\ifthenelse{\boolean{showcomments}}
{\renewcommand{\nb}[2]{
        \fbox{\bfseries\sffamily\scriptsize#1}
        {\sf\small$\blacktriangleright$\textit{#2}$\blacktriangleleft$}
    }
    
}
{\renewcommand{\nb}[2]{}
    
}

\newcommand{\ie}{\textit{i.e.},\xspace}
\newcommand{\eg}{\textit{e.g.},\xspace}
\newcommand{\etc}{\textit{etc.}\xspace}

\newcommand{\ellipsis}{\textit{…}}

\newcommand{\resp}[1]{$R_{#1}$}
\newcommand{\iresp}[1]{$RI_{#1}$}

\newcounter{findingcounter}
\newtotcounter{totalfindings}
\ifthenelse{\boolean{showboxes}}
{
    \newcommand{\finding}[1]{%
      \refstepcounter{findingcounter}
      \begin{tcolorbox}[boxsep=1pt,left=2pt,right=2pt,top=1pt,bottom=1pt]%
      \small
      \textbf{Finding \arabic{findingcounter}:} #1
      \end{tcolorbox}%
      \addtocounter{totalfindings}{1}
    }
    
  }
{
    \newcommand{\finding}[1]{}
}

\newcounter{discussioncounter}
\newtotcounter{totaldiscussion}
\newcommand{\discussionheading}[1]{%
  \refstepcounter{discussioncounter}
    {\textbf{\ref*{sec:discussion}.\arabic{discussioncounter} \hspace{3pt} {#1}.}}
  \addtocounter{totaldiscussion}{1}
}

\newcounter{rwcounter}
\newcommand{\rwheading}[1]{
  \refstepcounter{rwcounter}
    {\textbf{\ref*{sec:background}.\arabic{rwcounter} \hspace{3pt} {#1}.}}
}

\newboolean{normalheadings}
\setboolean{normalheadings}{true}
\ifthenelse{\boolean{normalheadings}}
{
    
}
{
    \newcounter{demcounter}
    
}

\begin{document}

\begin{abstract}

Although tension between university curricula and industry expectations has existed in some form for decades, the rapid integration of generative AI (GenAI) tools into software development has recently widened the gap between the two domains. To better understand this disconnect, we surveyed 51 industry practitioners (software developers, technical leads, upper management, \etc) and conducted 11 follow-up interviews focused on hiring practices, required job skills, perceived shortcomings in university curricula, and views on how university learning outcomes can be improved. Our results suggest that GenAI creates demand for new skills (\eg prompting and output evaluation), while strengthening the importance of soft-skills (\eg problem solving and critical thinking) and traditional competencies (\eg architecture design and debugging). We synthesize these findings into actionable recommendations for academia (\eg how to incorporate GenAI into curricula and evaluation redesign). Our work offers empirical guidance to help educators prepare students for modern software engineering environments.

\end{abstract}

\maketitle

\section{Introduction}
\label{sec:intro}

In recent years, generative artificial intelligence (GenAI) technologies have quickly changed the way software development is performed \cite{acharya2025generativeaitransformationsoftware}. Tools based on large language models (LLMs) such as GitHub Copilot \cite{githubcopilot}, OpenAI Codex \cite{codex}, Claude Code \cite{claudecode}, and Windsurf \cite{windsurf} provide developers with information retrieval, code generation, and more directly within their IDEs. Despite these rapid advancements, the ways in which we prepare aspiring software developers have not moved as fast \cite{ariza2025scoping}. Although industry is adopting the latest technologies and looking for candidates capable of utilizing them well \cite{gröpler2025futuregenerativeaisoftware}, more work needs to be done to understand the best ways to teach software engineering (SE) in the age of GenAI.

In many ways, GenAI has upended traditional education \cite{jie9}, including computer science (CS) curricula \cite{ariza2025scoping}. Current GenAI tools' code generation capabilities allow completion of certain introductory-level assignments with a single prompt \cite{10554754}, raising questions about how to redesign such assignments and ensure students learn foundational programming skills. The efficacy of these tools has also raised questions regarding which skills are becoming more important for development tasks, and which may see diminished relevance in the coming years. Despite these uncertainties, the industry continues to adopt GenAI tools at a rapid pace, and educational institutions are struggling to keep up.

As industry GenAI adoption accelerates \cite{McKinsey2024}, researchers have begun investigating GenAI in education. Existing literature reviews suggest heightened needs for strong problem-solving and soft skills like adaptability and communication \cite{necula2023artificialintelligenceimpactlabour,smuts2022society}. Despite concerns about academic integrity \cite{yurtseven2025education}, poor AI detection performance \cite{10554754} necessitates integrating GenAI into school curricula. However, historical disconnects between academia and industry \cite{scott2002skillgap, scott2004testing} suggest that care is needed when making these adjustments. While previous work has explored the question from an academic perspective, there has been an overall lack of industry input on how to best adapt current CS/SE curricula to the age of GenAI. There have also been no investigations into what hiring decision-makers will expect from new graduates as they enter the workforce. As such, essential questions remain unanswered, including which skills the industry values, how hiring practices have changed, the limitations that are perceived in academia, and what industry believes can be done to best prepare the next generation of software engineers. %

In this study, we present a novel investigation consisting of an analysis of 51 survey responses and 11 follow-up interviews with industry practitioners to uncover the qualities they are searching for in new SE hires and inform decisions about how to prepare aspiring software engineers for this rapidly evolving field. 

This paper presents the following main contributions:
\begin{itemize}
    \item Investigation of the skills that industry practitioners expect from new SE hires in the GenAI era, based on 51 survey responses and 11 interviews.
    \item Analysis of personal and organizational training practices to enhance knowledge and abilities related to GenAI-powered SE.
    \item Identification of the shortcomings in current CS curricula from the perspective of those working with and in charge of hiring new graduates.
    \item Actionable recommendations for how CS/SE educational practices should evolve, including when, where, and how to integrate GenAI tools.
\end{itemize}

\section{Related Work}
\label{sec:background}

\indent \rwheading{Misalignment Between Academia and Industry} The misalignment between university curricula and industry expectations in CS/SE has plagued the field for decades~\cite{scott2002skillgap,scott2004testing}. Broad surveys of embedded software practitioners \cite{akdur2021embedded}, product managers \cite{estes2025product}, and across SE roles \cite{akdur2022skillgap, CICO2021110736} suggest a need for greater focus on soft skills, while other studies attempt to address the skills gap through automated job posting analysis \cite{chumwatana2025bridging, jaiswal2024ukskill} or developing a work-based degree~\cite{10.1145/3351287.3351292}. Much of this existing research predates the GenAI era or lacks mention of GenAI skills: we build on this work by surveying and interviewing software practitioners with direct experience using GenAI across industry about specific shortcomings of academia. \Cref{tab:related_work} compares the topics covered by this paper with other work seeking to understand the gaps between industry and education.

\rwheading{GenAI Usage in Industry} While a mid-2024 McKinsey survey found that 65\% of respondents' organizations were regularly using GenAI \cite{McKinsey2024}, 62\% were still in the piloting or experimenting phase by the end of 2025 rather than having GenAI fully integrated and scaled across organizations \cite{McKinsey2025}. A cross-industry analysis of ongoing workplace integration suggests the need for greater technical proficiency and adaptability \cite{admsci14060127}, while literature reviews of future skills for SE specifically point toward problem-solving and interpersonal abilities \cite{smuts2022society, necula2023artificialintelligenceimpactlabour}. However, much of the research surrounding the integration of GenAI for SE into industry is still largely theoretical literature reviews \cite{alenezi2025review, Ajiga2024EnhancingSD}, do not focus more on practices than opinions about education \cite{otten2025prompting}, and remaining critical challenges \cite{gu2025challengespathsaisoftware} make it very difficult to accurately predict the professional skills which will be most valuable in the future. Chen et al. \cite{chen_huo_nam_peruma_port_2025} surveyed 32 hiring professionals, corroborating our own findings that many organizations have not adapted their candidate evaluation processes to account for GenAI tools, but practicing engineers provide a different perspective than recruiters. To address these gaps, our study focuses on the necessities and opinions of practitioners created by GenAI integration, which can then be used as a basis to further adapt to future changes.

\rwheading{Academic Responses to GenAI} Educational institutions have begun responding to GenAI through various approaches. Initial reactions from teachers emphasized prohibition and academic integrity concerns \cite{ariza2025scoping, alam2024education, kadan2024investigation, yurtseven2025education, weber2024ai, nguyen2023ai}, though the poor performance of GenAI detectors (especially for code) \cite{10554754, weber2023testing, 10.1145/3639474.3640068, sadasivan2025aigeneratedtextreliablydetected, Walters+2023} raises the possibility of false cheating accusations or forbidden AI usage going unnoticed. This, along with the increasing importance of GenAI in the workplace, forces educational institutions to integrate GenAI into curricula rather than completely prohibit it. A recent study of how students' trust in GitHub Copilot evolved through extended use found that students were likely to mention that Copilot requires a competent programmer, especially after their trust in the tool decreased \cite{shah2025evolutionprogrammerstrustgenerative}. However, 57\% of students whose trust increased mentioned Copilot's correct code output~\cite{shah2025evolutionprogrammerstrustgenerative}, and this efficiency risks ``blind acceptance'' of poor-quality code \cite{acharya2025generativeaitransformationsoftware}. Li et al. \cite{10685663} and Sah et al. \cite{10663054} proposed frameworks and reviewed approaches respectively for integrating GenAI into curricula, but both rely primarily on theory without direct practitioner input. The lack of input from industry professionals in curriculum redesign efforts remains a critical gap, which we aim to address with our study. 

\begin{table*}[t]
\centering
\caption{Comparison of related work investigating GenAI, SE skills, and curriculum alignment.
\textbf{Methods}: SV = Survey, IV = Interview, LR = Literature Review, SR = Scoping Review, SM = Systematic Mapping.
\textbf{Focus}: I = Industry, A = Academia, I/A = Both. %
Checkmarks (\checkmark) indicate topics directly addressed; dashes (--) indicate topics not addressed.}
\label{tab:related_work}
\resizebox{\textwidth}{!}{%
\begin{tabular}{lcccccccccc}
\toprule
\textbf{Study} & \textbf{Year} & \textbf{Method} & \textbf{Focus} & \textbf{GenAI} & \textbf{Skills Gap} & \textbf{Hiring/} & \textbf{Curriculum} & \textbf{Soft} & \textbf{Practitioner} \\
 & & & & \textbf{Subject} & \textbf{Analysis} & \textbf{Expectations} & \textbf{Recommendations} & \textbf{Skills} & \textbf{Survey ($n$)} \\
\midrule
Our Study              & 2026 & SV+IV & I & \checkmark & \checkmark & \checkmark & \checkmark & \checkmark & 51  \\
Chen et al.~\cite{chen_huo_nam_peruma_port_2025}            & 2025 & SV    & I & \checkmark & \checkmark & \checkmark & \checkmark & \checkmark & 32  \\
Otten et al.~\cite{otten2025prompting}           & 2025 & SV    & I & \checkmark & --         & --         & \checkmark & --         & 72  \\
Estes et al.~\cite{estes2025product}           & 2025 & SV    & I & --         & \checkmark & \checkmark & \checkmark & \checkmark & 122 \\
Akdur~\cite{akdur2022skillgap}                  & 2022 & SV    & I/A & --         & \checkmark & --         & \checkmark & \checkmark & 628 \\
Ariza et al.~\cite{ariza2025scoping} & 2025 & SR    & A & \checkmark & --         & --         & \checkmark & --         & --  \\
Necula~\cite{necula2023artificialintelligenceimpactlabour}                & 2023 & SR    & I & \checkmark & \checkmark & --         & --         & \checkmark & --  \\
Smuts \& Smuts~\cite{smuts2022society}         & 2022 & LR    & I/A & --         & \checkmark & --         & --         & \checkmark & --  \\
Cico et al.~\cite{CICO2021110736}             & 2021 & SM    & I/A & --         & \checkmark & --         & \checkmark & --         & --  \\
\bottomrule
\end{tabular}%
}
\end{table*}

\looseness=-1

\section{Study Methodology}
\label{sec:methodology}

In this study, we investigate how the adoption of GenAI in the SE industry is changing developer workflows and the skills expected of new hires, with the goal of informing improved CS/SE curricula.

Our study addresses four research questions (RQs):
\begin{enumerate}[label=\textbf{\labelitemi \space RQ$_\arabic*$:}, ref=\textbf{RQ$_\arabic*$}, wide, labelindent=5pt, leftmargin=5pt]\setlength{\itemsep}{0.2em}
    \item \label{rq:1} \textit{What job skills are expected of new SE hires?} To establish what industry values in entry-level developers.
    \item \label{rq:2} \textit{How are developers and organizations preparing to work with GenAI?} To understand how GenAI is adopted in practice, including individual learning strategies and organizational training.
    \item \label{rq:3} \textit{What do industry practitioners perceive as shortcomings in current CS/SE curricula?} To identify where academic preparation diverges most sharply from industry needs in the GenAI era.
    \item \label{rq:4} \textit{How do industry practitioners believe CS/SE curricula should adapt to GenAI?} To translate observed gaps and workflow changes into concrete, practitioner-informed curriculum guidance.
\end{enumerate}

To answer these research questions, we conducted a survey of industry software engineers with experience using GenAI to gather an overview of practices and opinions, then followed up with interviews to gain deeper insight into specific issues and responses.

\begin{figure}
    \centering
    \includegraphics[width=\linewidth]{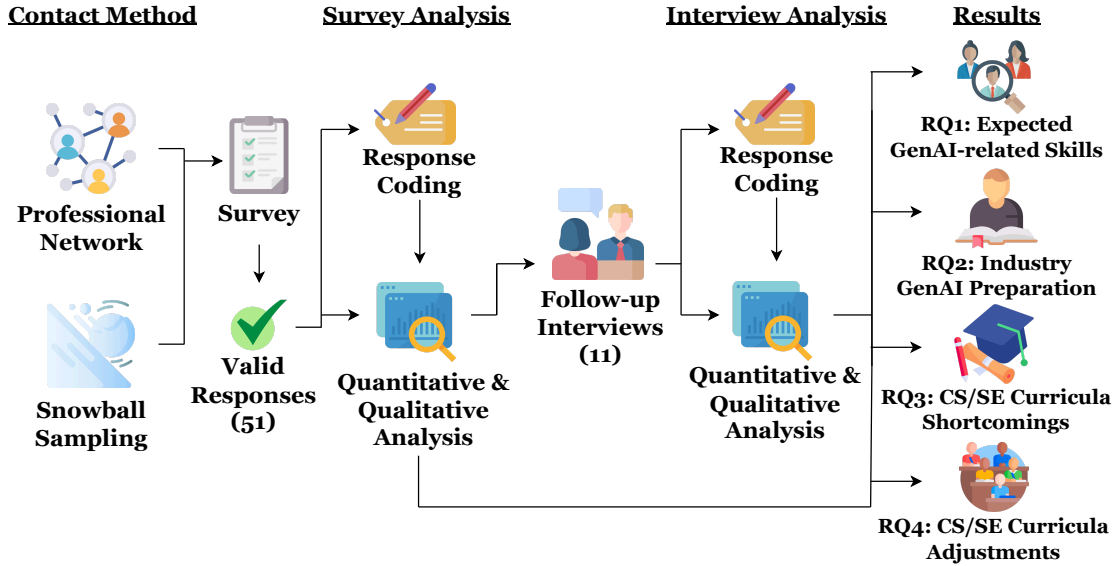}
    \caption{Research methodology (image credits at \cite{anonymous_repo})}%
    \label{fig:methods}
\end{figure}

\subsection{Participant Recruitment}

Our recruitment strategy targeted developers in industry with diverse backgrounds and experience levels. For survey distribution, we mainly relied on our professional networks and snowball sampling (\eg participants sharing the survey within their networks) by directly contacting potential participants via email or posting participation calls on professional social media platforms (\eg LinkedIn). Through this approach we were able to rapidly reach industry practitioners with recent GenAI experience with diverse backgrounds (see \Cref{demographics}). We decided against more open distribution channels without vetting participants, as this would have potentially allowed individuals without industry experience to participate in our study.

\subsection{Survey Design and Analysis}

Survey data informs RQ1, RQ2, and RQ3 through quantitative distributions of multiple-choice answers and open-ended questions. Our survey was organized into five sections, designed to follow general~\cite{survey} and SE-specific guidelines for survey design~\cite{DBLP:journals/sigsoft/PfleegerK01,DBLP:journals/sigsoft/KitchenhamP02,DBLP:journals/sigsoft/KitchenhamP02a,DBLP:journals/sigsoft/KitchenhamP02b,DBLP:journals/sigsoft/KitchenhamP02c,DBLP:journals/sigsoft/KitchenhamP03,linaaker2015guidelines,molleri2020empirically}. Questions were created and refined by all authors to balance survey length and clarity, mitigate bias, and ensure comprehensiveness. We used multiple formats depending on the type of information sought, including multi-select, Likert scales, and text fields. Branching logic was used to display certain follow-ups only to relevant respondents, such as those involved in the hiring process. The full survey text and logic is available in our replication packages ~\cite{anonymous_repo}. Sections 1 through 5 primarily establish the context necessary to answer RQ1 and RQ2, while sections 6 and 7 assist the interviews in answering RQ3 and RQ4.

\textbf{Section 1: Demographics} collects background information relevant to all of our RQs, allowing us to sort respondents and activate branching survey logic. Captured information includes respondents’ country of residence, years of software development experience and at their current organization, education level, primary organizational role, industry (derived from NAICS \cite{}), team size, hiring involvement, and organizational policy about GenAI tool use. The final question asks about GenAI usage frequency (Daily, Weekly, Monthly, Rarely, Never) for software development across five AI tool types (Chatbot web interfaces, IDE plugins, AI agents, Command line tools, and API integrations). Participants who selected ``Never’’ for all tool types ended the survey and were thanked for their time, ensuring that responses to subsequent questions reflect opinions developed through experience with GenAI.

\textbf{Section 2: Hiring Practices} was only displayed to respondents who did not select the “I have not been involved in hiring” option in Section 1. The first question asks how many candidates the respondent has evaluated in the previous twelve months. This is followed by two multiple choice questions asking about the preparedness level of recent university graduates for using GenAI professionally and whether the organization assesses GenAI skills during hiring. The section concludes with two open-ended questions about which GenAI skills are sought in candidates and how the evaluation process has changed due to GenAI.

\textbf{Section 3: GenAI Usage} was shown to all respondents for the purpose of gaining a deeper understanding about their opinions of and experiences with GenAI. It began with an open-ended question asking participants to list all the GenAI tools they use, followed by four Likert scale (Strongly Agree to Strongly Disagree) questions about whether participants find GenAI tools unreliable, can get desired GenAI outputs with time, consider themselves proficient GenAI users, and use GenAI more effectively than the average developer. The section ends with two multi-select questions asking how participants learned to use GenAI and how they stay up to date, with significant overlap in the available answer sets.

\textbf{Section 4: Organizational Expectations} contains only three questions intended to serve as a basis for follow-up questions and survey logic later on. The first two are Yes/No questions (with a third ``Unsure’’ option). The first asks whether the participant’s organization provides education about using GenAI tools, while the second asks whether the participant’s organization expects employees to use GenAI in daily workflows. The final multi-select question asks about organizational approaches to legal issues arising from GenAI outputs, such as whether this is a legal review process.

\textbf{Section 5: GenAI Education and Guidelines} featured questions which appeared in this based on participants’ answers in the previous section. If a participant indicated their organization provides GenAI education, they are presented with a multi-select question about the types of education (\eg training, guidelines, forums, \etc) and a Likert scale agreement question about the education’s sufficiency for preparing new hires. If a participant indicated their organization has specific GenAI usage guidelines, they are asked to describe such guidelines in an open-ended question. If a participant indicated their organization has specific GenAI usage guidelines, or that usage is discouraged or prohibited, they are asked what challenges these policies have caused. If a participant indicated their organization’s guidelines are unclear or non-existent, they are asked what the lack of policies has caused.

\textbf{Section 6: Follow-Ups} featured five open-ended questions. The first question participants were shown was dependent on the previous section’s Likert scale agreement question about the education’s sufficiency. If the participant agreed, they were asked how the education succeeds; if they were neutral or disagreed, they were asked how the education is insufficient. The following four questions were presented to all participants, asking what they wished their teammates knew about GenAI, which traditional software engineering skills GenAI has made more and less important, and which GenAI skills are most important for software development success.

\textbf{Section 7: University Offerings} gathered participants’ opinions about GenAI-related practices throughout universities and education. Participants were asked to rate their familiarity with university curricula in Software Engineering or related technical fields as either Very, Somewhat, Slightly, or Not familiar, with a brief description of each and a final ``Not applicable’’ option. The survey concluded with two open ended questions asking about the shortcomings of university curricula and advice for junior developers to use GenAI. A final optional text field allowed participants interested in a follow-up interview to provide their email address for further communication.

We collected 51 survey responses using Qualtrics~\cite{qualtrics} between 1 Nov. and 31 Dec. 2025. To analyze responses, we employed a mixed methods approach, combining quantitative and qualitative techniques. For multiple choice and Likert scale questions, we calculated data distributions and descriptive statistics to identify trends and patterns.

Three annotators (\ie authors) were involved in the analysis of open-ended text responses. Each response was assigned to two of the three annotators for independent \textit{open-coding} by choosing one or more \textit{codes} to represent the response using a shared codebook (\ie a spreadsheet). ~\cite{spencer2009card}. The third author, unassigned to that response, served as a tiebreaker when the two primary annotators’ codes did not align and a resolution could not be reached. Disagreements arose infrequently, typically from ambiguous responses that warranted multiple codes. Through this process, the annotators resolved any discrepancies, which included differing interpretations of potentially ambiguous responses, ensuring the same codes were applied consistently in different contexts, and using broader codes rather than more specific ones. We do not report inter-rater agreements because the list of possible codes grew and evolved throughout the process, making true agreement levels a poor metric that would be difficult to accurately calculate.

Ultimately, this process resulted in a final set of codes, definitions, and labels for each response,  found in our replication package~\cite{anonymous_repo}. Since codes were developed iteratively through an inductive process without having an initial codebook and multiple codes could be assigned to each response, we did not base our analysis on inter-annotator agreements. We followed best open-coding practices \cite{spencer2009card} and used annotator discussions to ensure the reliability of the results. When presented in this study, quotations from open-ended questions have been corrected for grammar and spelling, with ellipses or brackets indicating any text deletions or additions made for clarity or space-related concerns. For traceability, responses are attributed to participants using anonymous IDs (\eg \resp{12}).
\looseness=-1

\vspace{-0.1in}

\subsection{Interview Design and Analysis}

We conducted semi-structured interviews to gain deeper insight into survey responses. Interview data informs all RQs, especially on RQ3 and RQ4 where nuanced practitioner reasoning is most important. Each interview included 6-12 core questions adapted from a common protocol, with follow-up questions tailored to individual responses. These questions are found in our replication package~\cite{anonymous_repo}.
This flexible interview process allowed us to gather data consistently across a range of topics while adapting to focus on key ideas as they appeared.

We contacted every respondent (26/51) who included contact information at the end of the survey. Of these, eleven responded and scheduled interviews, which were conducted via Zoom~\cite{zoom} between 19 November and 19 December 2025, and were recorded using Zoom's built-in recording functionality. Interviews lasted 30-60 minutes and followed best practices to ensure participant safety and privacy \cite{kaiser2009protecting}. %
Upon completion, the recordings were transcribed with OpenAI's whisper-base model~\cite{whisper}. %
Three authors manually corrected these transcripts by comparing against the recordings. After verifying transcript accuracy, the authors iteratively identified and grouped key responses into thematic categories. The classifications of all quotations were reviewed and agreed upon by all three authors. When reporting quotations throughout this study, we have omitted filler words (\eg ``like'', ``um'') and indicated edits made for clarity with brackets and ellipses. To preserve the anonymity of our interviewees, we assigned each interviewee an identifier to attribute quotations (\eg \iresp{1}). %

\subsection{Participant Demographics}
\label{demographics}

Our survey was completed by 51 participants from eleven countries, spanning four continents: North America (61\%), Asia (25\%), Europe (12\%), and Oceania (2\%). %
Our participant pool showed a balance of academic experience with 35\% respondents having obtained a Bachelor's degree, 37\% a Master's degree and 27\% a Doctorate. %

Years of professional development experience ranged from 0 to 24 (mean = 8.45, median = 8): 23 had 10+ years of experience and two respondents reported no experience. %
(The industry roles of these two respondents, \ie Principal Data Scientist and Project Technical Lead, may not be directly aligned with software development. However, their responses to other survey questions indicate that they possess relevant hiring and industry experience related to GenAI for SE. For these reasons, their responses were included in the analysis.)
35\% of respondents indicated that they had fewer than five years of development experience, which might suggest that they are recent graduates with a current perspective on university curricula. The remaining 65\% reported five or more years of development experience, strongly suggesting that they are familiar with the skills required for day-to-day development work.

Respondents reported working in their current organization for fewer years on average than their total development experience (mean = 3.6, median = 3). %
Most were relatively new hires (76\%), having worked at their respective organizations for four or fewer years. There were only three outliers (6\%) with 10+ years at their current organization. 
This might reflect job churn in the SE industry or modern employment practices,  %
but also shows that many respondents recently went through hiring processes themselves and that others have been in their organizations long enough to see adaptations to new technologies.

We asked respondents to identify which of twenty industrial sectors best described their organization 
(as defined by NAICS~\cite{naics}). Ten sectors were represented. The most common were Information (computing, data-processing, web hosting, telecommunications, media, \etc) (18), Professional, Scientific and Technical Services (legal services, accounting, engineering services, \etc) (13), Health Care and Social Assistance (7), and Finance and Insurance (5). %

When asked for their primary role, nearly a third of respondents (16) selected Programmer/Developer.  Other common responses included Researcher (9), and Project Technical Lead (6), Educator (3), and Upper Management (3). Other roles with two or fewer occurrences included software architect, data scientist, analyst, applied scientist, DevOps/MLOps engineer, IT manager, consultant, security engineer, tester/QA engineer, and project manager. 
This diversity of roles likely also provides a diversity of perspectives on how GenAI is shaping industry and the future of academia.

Participants reported experience with a wide range of GenAI tooling: Chatbot interfaces (94\%), IDE plugins (86\%), API integration (73\%), AI Agents (67\%), and GenAI command-line tools (65\%). Most respondents indicated using IDE plugins (63\%) and web interfaces (61\%) daily. Participants also used a variety of model families, including ChatGPT (76\%), Claude (49\%), and Gemini (37\%). Most importantly, all respondents noted at least some experience with GenAI tools.  When asked about their self-perceived GenAI proficiency, 39\% strongly agreed that they were proficient, 35\% somewhat agreed, 20\% were neutral, and only 6\% somewhat disagreed.

Of the eleven interview participants, a majority were based in the United States (7), with two from Bangladesh and one each from China and New Zealand. Interviewees had development experience ranging from 1 to 22 years (mean = 8.0, median = 9), with over half (6/11) reporting five or more years of professional software development experience. Six have Bachelor's degrees, three hold a Master's, and two have Doctorates. Industry distributions are as follows: Finance/Insurance (3), Healthcare (2), Real Estate (1), and Professional, Scientific, and Technical Services (1). The remaining four work in Information, with one of them also working on Utilities and Construction. \iresp{1}, \iresp{2}, \iresp{6}, \iresp{8} and \iresp{10} are employed as programmers/developers, while \iresp{5} and \iresp{7} are project technical leads. \iresp{4} identifies as a researcher, and \iresp{11} is the chief technology officer (CTO) of their organization. Two participants utilized the ``Other'' category: \iresp{3} described themselves as a mix of architect, consultant, and researcher, while \iresp{9} is an applied scientist.

\section{Results}
\label{sec:results}

\subsection{\textbf{RQ$_1$}: Skills expected of new SE hires} 

We asked respondents involved in the hiring process at their organizations what skills they look for in new hires, with a particular focus on skills related to effective GenAI usage.

\subsubsection{Hiring Experience}
Of 51 respondents, 41 (80\%) indicated some involvement in their organization's hiring processes, including conducting technical interviews (31), onboarding hires (25), reviewing applications (24), designing interviews (18), conducting cultural interviews (11), or making final hiring decisions (8). Much of this experience was also recent. Of these 41 respondents, 71\% reported evaluating at least one candidate in the last 12 months. This suggests that respondents are familiar with organizational expectations for new hires, making them well-suited to answer questions about job skills, current practices, and candidates' shortcomings.

\subsubsection{Expected GenAI Skills}
\label{rq1:skills}

Most respondents (56\%) indicated that during the hiring process, GenAI skills were either not assessed (18) or were only assessed indirectly (5). An additional six (15\%) were unsure whether such an assessment had been added to their hiring process. When assessments were present, it was most commonly only for certain roles (22\%). Only three respondents (7\%) noted that GenAI skills are considered for applicants to all technical roles.

Twenty-three respondents indicated their organizations do not explicitly assess GenAI skills of new hires, though \iresp{1} clarified that while ``[GenAI skills might] not [be] an organizational goal. [...] it's going to be [a] hiring manager [decision].'' As such, it is critical to understand the skills that those involved in the hiring process %
prioritize, even if they are not concrete organizational policy. %

Accordingly, we asked respondents an optional question regarding which GenAI-related skills they look for during candidate evaluation, which received 30 responses. We show the most common answers in~\Cref{table:skills_hiring}, excluding answers noting that the respondent did not consider GenAI skills (5). We also asked all 51 respondents an open-ended question pertaining to which GenAI-related skills they believed were the most important for modern software development, the most popular answers to which are shown in~\Cref{table:skills_important}. All skills submitted, including those with two or fewer references, can be found in our replication package~\cite{anonymous_repo}. The results show that our respondents highly valued prompting skills both while evaluating candidates (12) and as a software development skill overall (20). Similarly, those who evaluated candidates sought to examine potential hires' knowledge of GenAI limitations (5), while the sample as a whole highly valued critical evaluation of AI outputs (16).

\begin{table}[]
\begin{tabular}{lc} \hline
\multicolumn{1}{c}{\textbf{Skill}} & \multicolumn{1}{c}{\textbf{Count}} \\ \hline
Prompt or context engineering & 12 \\ 
Knowledge of GenAI limitations & 5 \\ 
Experience with GenAI tools & 5 \\ 
Critically evaluating GenAI outputs & 5  \\ 
Supporting skills (\eg reasoning, problem solving, \etc) & 5\\ 
Open-mindedness & 4  \\ 
Proficiency with code generation & 4  \\
Knowledge of GenAI abilities & 3 \\ \hline
\end{tabular}
\caption{Top-8 GenAI skills assessed during hiring}
\label{table:skills_hiring}
\end{table}

\begin{table}[]
\begin{tabular}{lc} \hline
\multicolumn{1}{c}{\textbf{Skill}} & \multicolumn{1}{c}{\textbf{Count}} \\ \hline
Prompting & 20 \\ 
AI output analysis & 16 \\ 
Effective AI use & 9 \\ 
Debugging and troubleshooting & 6 \\ 
Knowledge of GenAI limitations & 5 \\ 
Creativity and experimentation & 4 \\ 
Ability to understand problems & 4 \\ 
Software architecture and design & 3 \\ 
Quickly identify/discern problematic code & 3 \\ 
Critical thinking & 3 \\ \hline
\end{tabular}
\caption{Top-10 Most important GenAI skills}
\label{table:skills_important}
\end{table}

We also asked the respondents what they wished other developers on their team knew about GenAI. 9/51 answered some version of ``N/A,'' which implies that those respondents were satisfied with their coworkers’ understanding of GenAI. Not all developers were satisfied, however, and some commented that they wished that coworkers had a better understanding of GenAI limitations (8), prompting fundamentals (7), and the risks of overreliance (4).  Others wished that coworkers knew more about GenAI output review (7) and testing (2). Still others wanted coworkers to see the potential of GenAI tools (5) and know that they could be useful despite limitations (2). We were also reminded of the learning curve for picking up GenAI (2) and that humans are still better equipped to complete certain tasks (2). The importance of understanding the problem being solved (2) and of the GenAI basics was also highlighted.

Each participant was asked what advice they would provide junior developers on how to effectively utilize GenAI, and the most commonly recommended path was hands-on use in diverse environments.  Respondents advised junior developers to experiment with GenAI tools (13), such as finding problems to solve with them (8) or increasing the frequency of use (6). These actions are likely to help junior devs and students gain a better understanding of GenAI (6) while recognizing its strengths and weaknesses (4). Respondents also advised junior developers to learn SE basics (9), prompt engineering (9), and effective GenAI output analysis (9). Other advice included learning how to use GenAI tools responsibly (5) and taking GenAI-related courses (3). Four respondents would encourage junior developers to use GenAI less, and three would caution them to use it as a guide, not a solution generator. Responses with two or fewer mentions can be found in our replication package ~\cite{anonymous_repo}.

In a follow-up interview, \iresp{10} synthesized these diverse requirements into a practical analogy, where operational proficiency is more important than deep technical knowledge: ``Different tools have different pros and cons and require different prompting skills, because I think the goal is to learn to prompt. [...] You don't need to know how a car works to use a car, right? That's how I see this. Yes, it's cool to understand at a high level what an LLM does, but for most people, I don't think that's relevant. They need to know how to drive, so: how to prompt, how to give it context, how to get what they want to get done, but also to have the understanding to know whatever it's spitting out is actually correct.''%

\subsubsection{Traditional SE Skills}
\label{sec:trad_skills}

\begin{table}[t]
\centering
\caption{Traditional SE skills perceived as more or less important in the GenAI era by at least 4 respondents (More Important: $n$=51, Less Important: $n$=46).}
\label{tab:traditional_skills}
\begin{tabular}{llc}
\toprule
\textbf{Direction} & \textbf{Skill} & \textbf{Count} \\
\midrule
\multirow{6}{*}{More important}
  & Software architecture / design  & 15 \\
  & Code review                     & 13 \\
  & Testing                         & 12 \\
  & Debugging                       & 9  \\
  & General Knowledge               & 4  \\
  & Problem Solving                 & 4  \\
\midrule
\multirow{7}{*}{Less important}
  & Programming language syntax     & 10 \\
  & Low-complexity coding           & 6  \\
  & Knowledge of frameworks / libs.  & 5  \\
  & Memorization                    & 5  \\
  & Prototyping                     & 4  \\
  & Documentation                     & 4  \\
  & Writing Code                     & 4  \\
\bottomrule
\end{tabular}
\end{table}

Although GenAI introduces new skill requirements, it simultaneously alters the importance of traditional SE skills. We asked respondents which traditional skills have become more or less important in the GenAI era. The most common answers are presented in \Cref{tab:traditional_skills}.

Skills with increased importance are broadly clustered around tasks that GenAI is believed to struggle with. Of 51 respondents, 15 indicated that software architecture/design is more important now, with some offering more specific ideas about design patterns, engineering best practices, and system modularity. The next three most important tasks focus on ensuring the quality of existing code, including code review (13), testing (12), and debugging (9). In a world where code is increasingly created by GenAI, the tasks shift to designing systems which guide GenAI and verifying the resulting output. \iresp{5} largely focuses on design and architecture, and believes that ``one of those things [which] would be most beneficial to people coming into the work field is to understand how to design something end to end.''

Expertise in easily automated tasks is becoming less valuable. The respondents noted that programming language syntax (10 of 46), memorization (5), and knowledge of frameworks and libraries (5) are traditional skills that are becoming less important. Although many participants identified testing and debugging as important skills, two indicated that these are becoming less important, suggesting that GenAI may increasingly be capable of automating these tasks. Many responses suggested that implementation is becoming less important, and that code writing (4) and specifically low-complexity code (6), prototyping (4), algorithms (2), and test case creation (2) were also mentioned to be lacking significance.

When participants were asked about GenAI skills needed by new hires, some referenced traditional development skills needed to support the GenAI usage. The participants noted that creativity (4) and understanding of the problem that the problem is trying to solve (4) are very important for the success of modern software, along with critical thinking (3) and adaptability (2).

\vspace{-0.1in}

\subsection{\textbf{RQ$_2$}: Industry preparation for GenAI} 

We asked respondents how they are personally learning and staying up-to-date with GenAI tools, as well as how their organizations are preparing developers to use GenAI effectively.

\subsubsection{Self-Learning}

Given rapid  %
GenAI advancements, learning the latest developments is crucial to stay on the cutting edge. 

As part of a multi-select question, we asked our 51 participants how they initially learned to use GenAI (providing 13 options). %
Nearly all respondents (48/51, 94\%) indicated that they learned GenAI tools and techniques through experimentation (\ie they were self-taught).  Roughly half (26/51, 51\%) learned from peers or from official documentation/tutorials (21/51, 41\%).
These were followed by reading tech blogs/articles (19), YouTube/video content (16), social media (16), developer forums (14), code samples (14), structured online courses (9), conference talks/webinars (6), academic courses (5), and books/technical publications (4). Only one respondent indicated that they were just starting to learn GenAI tooling. %

In a similar question, we asked our 51 participants how they stay up-to-date with GenAI developments.
The most popular sources of information were social networks (27), tech articles (25), conversations with colleagues/peers (25), and video content (20).  These were followed by research articles (15), employer-provided materials (14), AI tool release notes (13),  online forums (12), conferences (9), professional networking groups (7), newsletters (7), academic courses (3), and mainstream news outlets (1). Only 4 indicated that they do not actively stay up to date with GenAI developments.

\subsubsection{Organizational GenAI Training}

Thirty-six participants (71\%) indicated that their organization provided some form of GenAI training or resources.  These respondents  
were presented with a follow-up question asking them to select all types of provided resources, with seven options in addition to a text field. (The text field was utilized by only one participant to describe a competition which took place during non-work hours.) The participants selected a mean of 3.9 options, showing that the majority of organizations with GenAI training programs feature  diversity in their methods.

Organizations most commonly provided Formal training/workshops (26), Internal best practices (26), Online courses (24), Written guidelines (22), and Community forums (20), with External certifications (7) and Mentoring (5) being less common. The most common combinations of approaches were Formal training + Internal best practices (19), Formal training + Online courses (18), Written guidelines + Internal best practices (18), and Internal best practices + Community forums (18).

These same 36 participants were asked to rate their agreement with the statement ``My organization’s training and educational materials are sufficient for preparing new hires for development work in the era of GenAI.'' on a 5-point Likert scale. The majority (23) agreed, though only four of these were strong agreements. Seven somewhat disagreed and two strongly disagreed, while four remained neutral. We asked those who agreed how their organization's training succeeds, while the rest were asked where it fails. Common explanations of success mentioned that the training materials successfully increased GenAI skill (7) or adoption (4), provided useful advice (4) or practical experience (2), or were simple and easy to understand (3). \iresp{11}'s training tried to address all of these: ``[W]e gave general overview courses that talk[ed] on data security, some of the tools’ basics, the difference between the models, the personalities each of the models have, and then some `Prompting 101'-type basics.'' 

Examples of failure emphasized that the training was too narrow~(4), generic (3), or outdated (2). However, interviews revealed a separate set of shortcomings perpetuated by having model/tool providers take responsibility for organizational training. This approach can incentivize the provider to turn training into a sales pitch, rather than focusing on what is necessary to improve developers' skills. \iresp{8} portrayed this situation: ``[The AI vendor holds] office hours, where [...] people can come in and ask questions and figure out how to use these different tools more effectively. The problem is [that the representative] is not a very technical person. [...] They turn into marketing decks, [...] trying to get you to use the newest model.'' This willingness to outsource training to those with motives beyond providing safe, productive development environments could result in poor purchasing decisions and security vulnerabilities. As  \iresp{3} put it: 
``I accept that it sounds amazing, but my research also shows just how potentially vulnerable these MCP tools can be. How do you pass those tokens through without [authentication and authorization]? [I]t just doesn't feel very mature yet.' [E]ventually, they told me that no [similar institutions] had enabled MCP servers and yeah, I was right. It's not really secure enough for the type of environment that we work in yet.''

\subsection{\textbf{RQ$_3$}: Perceived  CS/SE curricula gaps}
We asked respondents about their views on the current CS/SE curriculum at the university level, with a particular focus on perceived weaknesses. We did this through an optional open-ended question which received 48 responses.

\subsubsection{Familiarity with curricula}
When considering only the respondents who answered our optional question (48), the majority (73\%) noted that they were at least familiar with current university CS/SE curricula.  14/48 respondents (29\%) indicated that they were very familiar, which means that they review curriculum, teach courses, or are involved in some way in curriculum development.  Only 12/48 respondents (25\%) claimed to have no or very limited familiarity with what is currently taught. A respondent (2\%) indicated that they had no formal university education.  When we consider years of experience, we observed that 35\% of respondents had fewer than five years of development experience, potentially suggesting that these respondents are  recent graduates who have more familiarity with current university practices.

\subsubsection{Insufficient Incorporation of GenAI} \label{rq3:insufficient_genai}
One of the most frequently mentioned shortcomings of current curricula was the limited coverage of topics related to GenAI (23\%). \resp{6} commented, ``university curricula fall short of teaching practical skills that are helpful in the workplace. [...] It's almost like there's a disconnect between what academics think software engineers will do in their day-to-day jobs and what they actually end up working on. [...] [U]niversities should be teaching how to work in tandem with GenAI - how to best leverage all the tooling available, their capabilities and limitations, [and] how to structure work around it.'' The sentiment that curricula adjust too slowly to new technology was explicitly mentioned by 5/51 respondents (10\%). \resp{43} summarizes: ``[S]chool curricul[a] struggle to keep up with industry demands.'' 4/51 respondents (8\%) expressed that they believed current curricula over-represent traditional CS/SE concepts, while neglecting newer advancements.

\resp{11} also raised that ``[s]ome universities prohibit [students] from using GenAI.'' \iresp{3} explained the impact of academia's apprehension to integrate GenAI into coursework: ``[I]t's a bit weird if at school we're like, `Oh, don't use it, because that's cheating,' and then [when] you get to the workplace [they say,] `You have to use it. Why are you not using it?' That would be quite jarring.'' \resp{13} reminds us that even when GenAI is included in the curriculum, it ``needs to be taught well [or s]tudents [will] end up using it as `write all my code [for me]', which ends up being [a] problem for everyone.'' In the same vein, respondents indicated that the limitations of GenAI (4\%) and its ethical/responsible usage (4\%) lacked coverage.

3/51 respondents (6\%) lamented the poor quality or absence of dedicated GenAI courses. \resp{10} reported that there are ``[n]o dedicated courses for GenAI'' and that schools are ``depending on years old curriculum that are virtually not followed [in industry] now.'' \resp{24} expressed a need for students to know how to develop software effectively in the modern era: ``They need to have a class regarding AI Software Development.''

\subsubsection{Impracticality of curriculum and assignments}
\label{rq3:impractical}
Equally important to the respondents was the perceived impracticality of the current university course work (23\%). Respondents commented on the ``high emphasis on how the systems work, with less material related to practical applications'' (\resp{32}) and 
how ``[e]xams and projects focus too much on rote learning rather than finding problems students would be interested in learning how to solve'' (\resp{33}). \iresp{1} noted that students ``are not being taught to solve problems with a heuristic approach of using all tooling available.'' Related to insufficient student preparation, \resp{54} observed that ``[s]tudents rarely build production-grade systems. [...] Industry tools (GitHub Actions, Terraform, Postman, Kafka, \etc) are often missing. [...] Graduates can write clean code but struggle to build, test, deploy, and maintain real systems.'' \iresp{5} came to a similar conclusion: ``[C]ollege graduates us[ing] stuff like git or unit testing [is] not commonplace. [T]here [are] a lot of general software engineering practices that [are]n't being implemented or [have] heavy emphasis on algorithms [instead of] solving business use cases.'' 
\iresp{6} echoed this thoery-vs-practice tension, noting that machine learning courses ``tend to teach [the] basics and the philosophies of these things, but if they actually want people to learn about the practical uses of these [tools], I think they should actually establish dedicated courses to generative AI.'' In short, it was not uncommon for respondents to believe that ``[u]niversities do not prepare students for the real world'' (\resp{42}).

\subsubsection{Other notable shortcomings}

Other shortcomings were mentioned less frequently, but remain potentially instructive. Respondents indicated a need for redesigned student evaluations and assignments. \resp{44} noted there is ``[n]ot enough time for SE. A curriculum with just one SE course will quickly fall behind.'' \resp{4} echoed this, wanting improvements in software testing and code review education, both increasingly important with influx of AI-generated code. \resp{15} also highlighted the increasing importance of ``interdisciplinary roles in tech (such as applied scientist, research scientist, AI engineer, AI SWEs)[, which all] have different scopes and goals and skills.'' In contrast to most other survey respondents, \resp{52} suggested that universities are being ``too reactive,'' likely referring to GenAI adoption occurring with limited foresight.

\subsubsection{Preparedness of Recent Graduates}
Similarly, we asked all participants (41) who had been involved in their organization's hiring process how well-prepared they believed recent university graduates were to use GenAI tools effectively in professional software development. Of these, 10/41 (24\%) indicated that they were uncertain or had no experience with recent graduates.  (We note that not all potential hires will be recent graduates.) In what follows, we consider only the 31 respondents with relevant experience. The full results can be seen in \Cref{tab:preparedness}.

20/31 respondents (65\%) indicated that recent graduates are somewhat prepared to use GenAI tools effectively. This indicates that there is still room for improvement, but that applicants are showing at least some skills which are important to employers.  However, based on follow-up interviews, it is possible that these applicants gained GenAI skills through self-paced learning or by working on side projects. Soft skills, such as critical thinking and problem solving, or traditional skills may also be transferred effectively to the usage of GenAI tools. Only one respondent indicated that recent graduates are very prepared. 5/41 respondents (16\%) reported that recent graduates were somewhat unprepared and three (10\%) reported that they were very unprepared, suggesting that more than a quarter of respondents had some dissatisfaction with the current university curricula. 2/41 (6\%) respondents believed that new graduates were neither prepared nor unprepared, which could also be seen as an indictment of the university system since GenAI skills are rapidly becoming central to development practices.

\begin{table}[t]
\centering
\caption{Industry practitioners' assessment of recent graduates' preparedness to use GenAI tools effectively in professional SE ($n$=31 respondents with relevant hiring experience).}
\label{tab:preparedness}
\begin{tabular}{lcc}
\toprule
\textbf{Preparedness Level} & \textbf{Count} & \textbf{\%} \\
\midrule
Very prepared        & 1  & 3\%  \\
Somewhat prepared    & 20 & 65\% \\
Neither / unsure     & 2  & 6\%  \\
Somewhat unprepared  & 5  & 16\% \\
Very unprepared      & 3  & 10\% \\
\bottomrule
\end{tabular}
\end{table}

\subsection{\textbf{RQ$_4$}: Adjustments to CS/SE curricula}

During follow-up interviews, we asked respondents their thoughts on how GenAI should be taught, with a particular focus on how the current CS/SE curriculum at the university level could be improved.

\subsubsection{Integrated GenAI vs. Standalone Courses}
\label{rq4:how-to-incorporate}
Interviewees presented a nuanced view of how they believed that GenAI could best be incorporated into university curricula, with several suggesting that GenAI should be woven throughout the existing curriculum rather than being added through a dedicated course. On this, \iresp{9} noted that GenAI ``should be included in software engineering subjects, [but] I don't think we need a separate course...'' while \iresp{11} cited that because ``[GenAI technologies are] so fundamental to the way that the practice is, it needs to be touched on almost everywhere.''

\iresp{4} further pointed out that a standalone course just about `How to Use GenAI' seemed like the type from an online learning platform. ``I don't really feel like we can easily design the structure of [that] course, because it might be very general [and] very hard to cover all the topics. People might [be] using an AI to write code, [to] write documents, or do a literature review. [T]hey are quite different [tasks].'' 
\looseness=-1

However, some interviewees valued dedicated GenAI courses. \iresp{6} mused that ``if [universities] actually want people to learn about the practical uses of [GenAI], [then] I think they should actually establish dedicated courses [...] teaching people how to use LangChain, chatbots, [and related concepts].''  \iresp{7} elaborated: ``Should students learn how to run a model or invoke APIs for [...] generative AI[...]? I think it should be an option for a course you can register for. It's certainly technical enough. [...] Those are skills you can sell.'' A dedicated course could potentially also cover the emerging security considerations outlined by \iresp{11}: ``There's a whole new slew [of] techniques that are needed to do secure-by-design with AI and agentic and semi-autonomous, really fully autonomous work. So data classification gets really important, hygiene in general, secrets hygiene, \etc gets really important. Understanding [how] to do attack models [...] prompt injection is extra scary now because of the autonomy that we're giving these systems.''

\subsubsection{When to integrate GenAI}
\label{rq4:when-to-integrate}

Some respondents also gave us their thoughts on when GenAI should be integrated into curricula. Interviewees broadly agreed on the importance of learning the fundamentals before relying heavily on GenAI.  \iresp{10} compares the situation to early math education: %
``You want to learn how to graph functions before you use a graphing calculator, right? I think this is the same.'' Furthering the analogy, \iresp{7} suggested that at some point, students are ready to move beyond the fundamentals: ``Just like we learn arithmetic by hand, even though we have calculators, you have to learn to think by hand without using GenAI. But we do use calculators [and] spreadsheets, we use so many things in educational environments today. When you learn accounting, it is no longer about the craft [of] add[ing] numbers together. It is about accounting. It is no longer about addition.'' 

In their survey responses, \resp{13} put it this way: ``I think for the core courses that teach you how to program (like Intro to Programming, [Object Oriented Programming], [Data Structures and Algorithms]) there should be no AI allowed since those establish your foundation. But for later courses where the idea shifts to domain knowledge/building things, there should be some leeway in its usage since people are go[ing to] be using it, esp[ecially] in the industry.'' \iresp{8} also noted the balance saying, ``there is a place for coursework that does not allow for the use of GenAI and a place where it should be encouraged. For example, [in] a[n] introduction to Python course, [... students] need to get in there, type the code, get used to the code because that's what builds that literacy. Once you're [in an advanced class] where you [already] know how to write a function, you know how to build custom libraries, and do [that] sort of thing, I don't think there's any harm at that point in using [GenAI].'' Lastly, as \iresp{2} pointed out, it is important that students understand \textit{why} GenAI usage is prohibited in certain instances: ``[S]tudents have to know `Hey, there's a balance between me learning and me being productive': a really significant balance, especially when it's the first few semesters or years.''

\subsubsection{Focus on Fundamentals}
\label{rq4:fundamentals}

As noted in the previous subsection, respondents believed that the fundamentals are still very important.  As \iresp{10} put it: ``I don't see AI as a tool that will replace [the fundamentals]. I see it as a tool that amplifies whatever you know.''  These fundamentals are particularly important for interacting with GenAI and evaluating its outputs.  \iresp{4} said that ``we need to encourage [students] to still learn basic concepts like data structure[s] [and] algorithm[s], and master them because they are central skill[s] for [...] verify[ing] if the AI generated content is correct and safe.'' There is a related need for universities to increase focus on good development practices such as architectural design, testing, code review, and maintenance. %
\iresp{8} described ``reviewing code, reading code, code literacy, [and] being able to look at complicated code and get a feel for what it does'' as ``exceptionally important.'' \iresp{4} cautioned us on the dangers of building technical debt: ``[Developers may] continue to develop without leaving notes or changes in the documents. So this leads to [a situation where] the final code is not always consistent with the documents. [...] I think this is the bottleneck for the agents to understand the repository %
[...] because [this information is] not specified anywhere where the agents can read [it] [...] [To] help the students to use LLMs better, we need to [teach] them to make [their] development clear and [avoid] leav[ing] a lot of tech debt in [their] projects, and also keep [a] record of all [their] historical design decisions.'' An even greater focus on gathering and formalizing requirements might be apt, as \iresp{11} saw ``prompting as a specification. [...] Nobody's enthused about requirement analysis [but] it's a necessary evil. So you kind of frame [the requirements and specifications] and really that's now at the prompt.'' %

Universities may also need to teach students to think like senior developers.  \iresp{5} commented that ``[t]here's a lot of stuff [where] you have to get very granular on the details, almost like you have a brand new junior developer that can work at a crazy pace. So I'm walking it through everything I want and all my expectations.'' \iresp{3} explained how ``a lot of companies feel like they don't need to hire a junior engineer anymore, because [...] if I just buy my senior a GitHub Copilot license then we don't need to hire that new junior, right?'' In this same vein, \iresp{10} also stressed that product understanding skills are becoming increasingly important: ``I think what's going to separate the engineers that [...] get left behind by GenAI, and the ones who use it to their advantage, is becoming more like product engineers where you're not just writing code, you're actively helping build the product on top of writing code.'' \iresp{10} noted that ``you have to do it the old fashioned way: read docs, ask people questions and absorb information. [...] I think that having conversations, asking questions and really understanding [...] how the business sells this stuff and how it fits into the greater product that the business is selling is very important.''

\subsubsection{Focus on Soft Skills}
\label{rq4:soft-skills}
Relatedly, respondents also suggested that soft skills such as problem solving, reasoning, and critical thinking are increasingly important in the age of GenAI. Two respondents explicitly indicated looking for reasoning abilities when evaluating candidates. \iresp{3} explains the importance of such skills: ``It's important for engineers to continue to use their critical judgment, [their] critical thinking to validate and verify all the code that [the model] generates because ultimately it's still their work and so they're accountable for the work that they commit.'' \iresp{11} put it this way: ``[Y]ou still need [theory and application] but it's more important that you understand the high level. The `how you solve these problems' is more important than how the problem is solved, more and more so. [...] [H]ow we solve problems hasn't fundamentally changed in decades. We have new tools that enable us to move faster in certain things that increase how quickly you can prototype. But the basics of how applications work, how humans interact with things. I think [this is] the most important thing: understanding the playbook of how you solve problems.'' The role of software developer may be changing, but it is not going away: ``the unit of work is [...] shifting from code to intent'' (\iresp{11}).

When evaluating candidates, \iresp{1} said, ``I'm more interested in people that are curious about the world around them, right? I want to know that you're aware of the tools, [that] you've played with them, you're interested in them. [...] Basically, can you solve problems?'' But unfortunately, as \iresp{2} comments, curiosity does not always survive the academic process: ``Curiosity is definitely one of [the ways current students are unprepared]. [P]eople go into CS because they're curious about computers. I think for some people that curiosity dies in CS when you actually get to programming.''

Academia can prepare students for the future by encouraging experimentation and fostering curiosity. \iresp{11} concludes with: ``I love experimentation, so, just as much as you can, encourage folks to double down, get in, and learn. Don't be afraid, just learn.'' %

\subsubsection{Focus on Practicality}
\label{rq4:practicality}
Four survey respondents (8\%) wrote that they believed university curricula focused too heavily on traditional concepts, and wanted to see changes to reflect how the field has changed. As \resp{17} put it, ``[...] universities still focus heavily on theory and traditional programming, and they haven't fully adapted to the fast pace of change in areas like GenAI.''

As noted in \Cref{rq3:impractical}, eleven (22.92\%) respondents also mentioned that current university curricula were impractical: that is, it does not sufficiently reflect the types of work that students would be asked to perform once they enter the workforce. Some interviewees, such as \iresp{1}, suggested remedying this by shifting to more real-world style learning: ``I think universities need to switch to a model [...] more like [the] trades. [I]n this case, where you're working and doing and you're basically training for the job you're going to do [...] maybe it's solving problems at university, maybe it's things like internships.'' Similarly, \resp{39} would advise junior developers to learn GenAI through ``contribut[ing] more to OSS,'' (open source software), and \iresp{11} wanted such contributions more formally encouraged: ``I think encouraging students to do open source contributions early on [in] their [CS] journey is important, because then you start to get the feel of how a lot of these commercial projects work\ellipsis.''
\looseness=-1

\subsubsection{Evaluations}
\label{rq4:evaluations}

Teaching GenAI may also require rethinking how students are assessed. Four (8.33\%) survey respondents explicitly noted that current curricula needed to redesign evaluations to accommodate the use of GenAI. Interviewees elaborated on changes they would like to see in academic evaluation. For example, \iresp{10} wanted to see a modern equivalent or revival of ``no-calculator exams [...] [I]f you do that, people will actually go and learn, right? Without using the tool.'' \iresp{8} also mused about a return to pen and paper examinations: ``[I]t's really difficult to use an LLM to your advantage when you have a paper and a pen.''

AI usage guidelines can extend beyond exams and feature in any assignment, as noted by \iresp{8}: ``[I]ndicating what the extent of the usage of AI tools should be or what is allowed for that project, that assignment, should be commonplace and standard occurrence.'' \iresp{8} went on to note that the if goal of assignments should be to demonstrate proficiency, this can be accomplished through a more thorough reading or interrogation of a student's work: ``Imagine you turned in a project [...] And you were allowed to use GenAI with the specification [that] you need to understand what code you turn in. [When the] professor receives those files, and instead of just grading them in a black box, [they schedule] 10 minutes with each team or student [...] And after they've looked through it, [ask,] `Tell me what this part of the code is doing.'\hspace{0pt}''

Perhaps the onus is on educators to thoughtfully design assignments that still teach students while they use GenAI. ``[Y]ou have to craft the questions in such a way where just asking ChatGPT or Claude Code doesn't answer your question. [T]here still needs to be thought behind it'' (\iresp{10}). \iresp{8} commented that assignments could be designed to reveal the limitations of GenAI: ``As an example assignment for students: Generate code that will do [something], then have the student identify problems with that code and provide a corrected version of the code.'' But with projects of sufficient size, \iresp{8} also noted that overreliance on GenAI isn't a realistic problem since ``[i]f [the students] try to vibe code a project and it's terrible and doesn't work, they're still going to learn because now they have to debug it or start over and write it from scratch anyways.''

\subsubsection{Other recommendations}
\label{rq4:other-recs}

Respondents provided several other recommendations that might be applicable to GenAI education. Many students may not have the resources to use state-of-the-art GenAI models, so \iresp{1} suggested that ``[GenAI tools] should be something that universities provide off the bat. Day one: here's your ChatGPT license. [...] It probably depends on the curriculum, but [...] they're in college. Man, how are [the students] going to afford that [on their own]?'' But institutions of learning may also face financial limitations, so \iresp{6} suggested potential partnerships ``with Coursera or Data Camp or platforms like that.''

\iresp{10} suggested multiple reasons for using older versions of models during instruction. First, they can be used to build prompting fundamentals as discussed in \Cref{rq1:skills}. %
They can teach adaptability in a changing landscape: ``[Y]ou could [suggest] `us[ing] this older version or this other tool,' and then [switching] to a newer one, and [the students] can go have that context switching.'' Using different tools can help students learn to assess and understand model capabilities: ``I think [the different tools] have different pros and cons, so maybe you have to pick the right tools to show `this tool is really good at this' [or] `this tool is really good at that'. [A]nd then [...] you hope students can contextualize the different knowledge and apply them across the tools.''

\iresp{2} advised continuous learning for faculty. %
``[E]very now and then, the faculty [should] work on a project where they have to code a software application that's not something very basic. [...] When [they] use [the AI tools], they're going to learn a lot more. [W]hen I say they're out of touch [...], it feels like a decent chunk of them [...] haven't even built a software project in a very, very long time. So they kind of don't know what they're talking about.''

Respondents also noted that GenAI is not just a challenge for universities, but that it also presents unique opportunities to improve existing courses and learning outcomes.  \iresp{10} ``if you're teaching [you can] prompt [the model] to give you a really good example of what [you] just talked about. So the theory is no longer theory [...] I think once there's code, [...] theory is no longer in their heads because you don't know what people are imagining. [The example] is concrete and in front of them. So that's where I think in school it can help. So, [...] in programming language[s], [...] it's nice to talk in theory, but maybe you can prompt [the model] like `show me an example' and then use that example to learn.''

Finally, \iresp{6} suggested that student organizations might be better positioned to fill the gap and provide students with GenAI experience as they ``[...] can have a greater impact on teaching students [GenAI], because they can be flexible. They can arrange a boot camp on the weekends and help [students] build a project on their own.''
\looseness=-1

\vspace{-0.1in}

\section{Actionable Recommendations for Education}
\label{sec:discussion}
We discuss the practical implications of our findings and offer ac-

\noindent tionable recommendations for CS/SE education.
\looseness=-1

\discussionheading{Encourage student buy-in}
As noted in \Cref{rq4:when-to-integrate}, there will be times when students' GenAI usage should be constrained or prohibited. %
This is particularly true for introductory courses where students learn the fundamentals.  In these cases, curricula must not only be redesigned around GenAI, but also be redesigned in a way that openly articulates any limitations imposed on students.  Students must understand that prohibitions on GenAI tools at this stage will facilitate the best learning outcomes.  Time should be dedicated to motivating the importance of fundamentals and demonstrating to students why they cannot simply rely on GenAI.  Without this grounding, it is a near certainty that students will circumvent guidelines and use GenAI tools in ways that boosts productivity but might replace critical thinking. Ultimately, finding this balance may become increasingly difficult as future students come to university with years of experience relying on GenAI tools in their daily lives and in their other courses.

\discussionheading{Focus on foundations}
As noted in \Cref{rq4:fundamentals}, fundamental skills such as algorithmic thinking, debugging, testing, and code review are becoming increasingly important with the emergence of GenAI tools in industry \cite{schmidt2025liberatinglogicageai}. Mastering these key concepts will help students design the best solutions, prompt models, and evaluate model output. The core principles behind SE remain, even if the position of \textit{coder} (like that of punch-card programmer before it) fades into memory.
In \Cref{{rq4:soft-skills}}, we saw the increasing importance of soft-skills in the CS/SE profession.  University CS/SE curricula should focus on teaching problem solving, reasoning, and critical thinking, (and perhaps this should be the case even in lower-level education). As \iresp{11} commented, ``[GenAI has moved] the unit of work [...] from code to intent.'' So schools should focus less on code-specific elements %
and more on the high-level abstractions and techniques that will continue to be relevant for as long as software is developed, evaluated, and used.

Universities should also be mindful when adjusting their criteria that such changes are not done only to chase the latest fad in GenAI.  There should be a balance between incorporating the latest-and-greatest tools, practical industry skills, and the fundamentals.  %
\resp{22} commented on this balance: ``[Universities] should not aim for hype technologies and [instead] keep to the foundations - that's done well and student[s] often criticize this. However maybe even the foundations should change now.'' So while it might be time to rethink SE curricula for the age of GenAI, universities should approach the topic thoughtfully and deliberately without making rash or poorly thought out decisions.

\discussionheading{Adjustments to assignments and evaluations}
The question of how to evaluate students while maintaining GenAI awareness is widely investigated in academia~\cite{combrinck2025student,ramezani2025redefining,lee2025socratic,hazzan2025rethinking}, in addition to being debated by large educational publishers~\cite{cengage} and news outlets~\cite{nyt-assess,nyt-ai-bad}. %
As shown in~\Cref{rq4:evaluations}, industry practitioners are no exception to this. 
One thought is to limit the ways in which students can use GenAI for assignments and evaluations, such as ``no-calculator'' or paper exams. Alternatively, educators can rework evaluations in a more fundamental way to ensure that learning goals are still met even when GenAI is allowed. The appropriate evaluation strategies may depend on course level: no-tool assessments suit foundational courses for early skill development, while tests redesigned to incorporate AI use are more appropriate in advanced courses focused on professional readiness. Our participants identified multiple ways to achieve this, including by carefully constructing GenAI usage guidelines for students or requiring them to be able to explain their work to the instructor.

The optimal approach may lie somewhere in the middle of these recommendations. %
Some practitioners also believed that GenAI belongs in certain parts of the CS/SE curriculum, but perhaps not all of it (\Cref{rq4:when-to-integrate}). A mixture of GenAI-enabled assignments and no-tool assignments could both promote students' understanding of how to use GenAI while also checking that they comprehend the underlying fundamentals that support SE. 
\looseness=-1

Considering GenAI's ability to generate code, assignments can be changed such that writing a basic program is no longer the primary objective. An approach is to raise the level at which students need to think about software: one of the benefits practitioners have identified from GenAI is the ability to focus on higher-level tasks such as software architecture and design~\cite{otten2025prompting}. Rather than requiring students to implement basic functionality, students might instead be provided with a basic implementation directly and asked to implement features, change algorithms, or make other adjustments in ways that require them to consider data structures and algorithms, adhere to certain constraints, or engage with other design-level tasks. As noted by practitioners in~\Cref{rq4:evaluations}, assignments might also require students to provide an explanation of what the code is doing or how it works, ensuring that even if the code was generated by GenAI, they still understand the core concepts required.

\discussionheading{Importance of practicality}
The impracticality in education was discussed by respondents (\Cref{rq3:impractical,rq4:practicality}), but this shortcoming may not be strictly related to the absence of GenAI in the curricula. Perhaps, the rapid changes in industry brought about by GenAI are the latest, and perhaps most prominent, manifestations of the tension between what is taught in universities and what is required for day-to-day development work. So, even beyond incorporating GenAI into curricula, it may be prudent for educators to address other foundational issues, such as the perceived impracticality in curricula.  Courses should teach practical skills, development practices, and industry-standard tools, all without neglecting theory and the fundamentals. Using GenAI, students should be encouraged to work on larger, real-world projects that are of interest to them, such as contributing to open-source repositories. In doing so, students can learn to read and modify existing code bases, apply good git practices, test solutions, participate in code reviews, and perhaps become part of a wider software community.

\discussionheading{GenAI courses}
As identified in \Cref{rq3:impractical}, %
current GenAI courses often focus on the underlying technology (\eg how machine learning and language models work) rather than how to apply it to SE. As reported in~\Cref{rq4:how-to-incorporate}, the interviewees had divided views on whether a dedicated GenAI course would be valuable in CS education. However, as~\Cref{rq3:impractical} and ~\Cref{rq4:practicality} make clear, the professionals we surveyed would prefer if university curricula dedicated more time to providing practical projects that are representative of what a modern SE position will entail. As noted by \iresp{1} in \Cref{rq4:other-recs}, universities should consider offering paid GenAI tool access to their students for academic usage. Universities should also ensure that students can easily determine which tools are available to them. 

From this we can derive two main conclusions. First, while the extent and means by which GenAI is integrated into coursework may vary according to the level of the course and the nature of work within it (see~\Cref{rq4:how-to-incorporate}), SE students should be taught how to apply GenAI to their discipline experientially. This could occur in a dedicated GenAI course, or it could be woven throughout existing courses. Second, courses that seek to teach CS students about GenAI could benefit from taking an applied approach: while a foundation of theory is important (\Cref{rq4:fundamentals}), students could benefit from an understanding of 1) effective prompting strategies for SE, 2) how GenAI models can fail, and 3) how to supervise and correct AI-generated code. We also suggest that dedicated courses cover secure and responsible usage of GenAI tooling. %

\section{Threats to Validity}
\label{sec:threats}

\textbf{Internal Validity.}
To mitigate possible coding bias, we used a rigorous iterative open-coding methodology to analyze open-ended survey responses. We also followed best practices when formulating our survey questions in order to avoid confusing and/or biasing language. 
The sample population was largely sourced from our professional networks with snowball sampling. Given that 27\% of respondents hold doctorates (a higher-than-average proportion of the broader software engineering workforce), our sample may have a close proximity to academia that could influence their perspectives on university curricula. However, familiarity with academia may be a strength, as it increases the likelihood that curriculum suggestions are grounded in direct experience. Additionally, there is the threat of self-selection bias, since the sample may skew toward practitioners who are enthusiastic about GenAI and willing to engage with surveys on the topic. Overall, self-reported familiarity with university curricula could be an unreliable indicator: recent graduates might have experienced only one institution, and GenAI’s rapid advancement means that perceptions and academic experiences will continue to evolve. Future work should seek greater recruitment across industries, experience levels, and educational backgrounds to validate our findings.

\textbf{External Validity.} In this study, we sought to understand and address the many challenges of integrating GenAI into SE university education to support industry practice. Though we attempted to  provide solutions in a way that could be valuable for all stakeholders, we must acknowledge that every institution has unique opportunities and limiting factors. As such, we intend our suggestions to be a starting point, with further research necessary to evaluate their effectiveness in a variety of environments. 

\textbf{Construct Validity.} This work relies on self-reported developers' perceptions, which can be subjective and influenced by recall and social desirability biases. We sought to mitigate this threat by anchoring subjective assessments with more concrete metrics where possible and by anonymizing results. The demographic information provided in our study was supplied by respondents and cannot be independently verified.

\section{Conclusion}
\label{sec:conc}

GenAI's rapid integration into software development demands urgent alignment between academic CS/SE  curricula and industry expectations. Through a survey of 51 practitioners and 11 follow-up interviews, we revealed industry perspectives largely underrepresented in empirical literature about curriculum design. Our findings suggest that new skills related to GenAI use (prompting, output analysis, \etc) are now essential. Our participants consistently emphasized that high-level skills like architectural design, testing, and code review are becoming more important in the GenAI era, while low-complexity coding and language/syntax/framework memorization seem to be losing favor. We provide actionable recommendations, such as disallowing AI usage until students master core competencies, integrating GenAI into existing classes, and redesigning assessments around either GenAI usage or prohibition. Future work should explore academic perspectives about implementation logistics and longitudinal effects on students.

\bibliographystyle{ACM-Reference-Format}
\bibliography{references}

\end{document}